 \documentclass[twocol]{ametsocV6.1}

\newcommand{\bsub}{\begin{subequations}}
	\newcommand{\esub}{\end{subequations}}

\newcommand{\ka}{\kappa}

\newcommand{\ep}{\epsilon}

\newcommand{\pat}{\partial}

\newcommand{\x}{\times}

\newcommand{\beq}{\begin{equation}}
\newcommand{\eeq}{\end{equation}}
\newcommand{\bsubeq}{\begin{subequations}}
	\newcommand{\esubeq}{\end{subequations}}
\newcommand{\beqn}{\begin{eqnarray}}
\newcommand{\eeqn}{\end{eqnarray}}
\newcommand{\fr}{\frac}
\newcommand{\lb}{\label}
\newcommand{\er}{\eqref}

\usepackage{lineno}

\usepackage{hyperref}
\hypersetup{colorlinks,allcolors=blue}
\usepackage{amsmath}
\usepackage{multirow}
\usepackage{amsfonts,amssymb}
\usepackage{graphicx}
\usepackage{subfigure}
\usepackage[rel]{overpic}

\usepackage[all]{background} 
\usepackage{url}
\SetBgContents{Published by Journal of the Atmospheric Sciences}
\SetBgColor{gray}
\SetBgPosition{6.5,-20}
\SetBgOpacity{0.75}
\SetBgAngle{0}
\SetBgScale{1.2}

\usepackage{tikz,xcolor,hyperref}
\definecolor{lime}{HTML}{A6CE39}
\DeclareRobustCommand{\orcidicon}{%
	\begin{tikzpicture}
	\draw[lime, fill=lime] (0,0)
	circle [radius=0.16]
	node[white] {{\fontfamily{qag}\selectfont \tiny ID}};
	\draw[white, fill=white] (-0.0625,0.095)
	circle [radius=0.007];
	\end{tikzpicture}
	\hspace{-2mm}
}
\foreach \x in {A, ..., Z}{%
	\expandafter\xdef\csname orcid\x\endcsname{\noexpand\href{https://orcid.org/\csname orcidauthor\x\endcsname}{\noexpand\orcidicon}}
}

\title{The mean wind and potential temperature flux profiles in convective boundary layers}

\authors{Luoqin Liu\orcidL{}\aff{a,b}\correspondingauthor{Luoqin Liu, luoqinliu@ustc.edu.cn}, 
Srinidhi N. Gadde\orcidS{}\aff{b},
Richard J.A.M. Stevens\orcidR{}\aff{b}
}

\affiliation{\aff{a}{Department of Modern Mechanics, University of Science and Technology of China, Hefei 230027, Anhui, China}\\
\aff{b}{Physics of Fluids Group, Max Planck Center Twente for Complex Fluid Dynamics, J. M. Burgers Center for Fluid Dynamics, University of Twente, P.O. Box 217, 7500 AE Enschede, The Netherlands}
}

\abstract{
We develop innovative analytical expressions for the mean wind and potential temperature flux profiles in convective boundary layers (CBLs). CBLs are frequently observed during daytime as the Earth's surface is warmed by solar radiation. Therefore, their modeling is relevant for weather forecasting, climate modeling, and wind energy applications. For CBLs in the convective-roll dominated regime, the mean velocity and potential temperature in the bulk region of the mixed layer are approximately uniform. We propose an analytical expression for the normalized potential temperature flux profile as a function of height, using a perturbation method approach in which we employ the horizontally homogeneous and quasi-stationary characteristics of the surface and inversion layers. The velocity profile in the mixed layer and the entrainment zone is constructed based on insights obtained from the proposed potential temperature flux profile and the convective logarithmic friction law. Combining this with the well-known Monin-Obukhov similarity theory allows us to capture the velocity profile over the entire boundary layer height. The proposed profiles agree excellently with large-eddy simulation results over the range of $-L/z_0 \in [3.6\times10^2, 0.7 \times 10^5]$, where $L$ is the Obukhov length and $z_0$ is the roughness length.
} 

\begin{document}

\maketitle

%
%
%
%
%

%

\section{Introduction} \lb{sec.introduction}
Convective boundary layers (CBLs) are frequently observed during daytime as the Earth's surface is warmed by solar radiation \citep{stu88}. Due to their frequent occurrence, the fundamental understanding of CBLs is highly relevant to agriculture, architectural design, aviation, climate modeling, weather prediction, and wind energy applications, to name a few. The modern scientific literature on CBLs goes back over 100 years. Initially, the focus was on low-altitude measurements, and with the introduction of more advanced measurement techniques, the focus gradually shifted upwards. However, only after the introduction of large eddy simulations (LES) in the early seventies, it has become widely accepted that thermodynamic indicators are most suitable to identify the different CBL regions \citep{lem19}. However, obtaining analytical profiles that describe the wind and potential temperature flux in the entire CBL has remained challenging due to the different flow physics in the various CBL regions.

\begin{figure*}[!tb]
\centering
\begin{overpic}[width=0.8\textwidth]{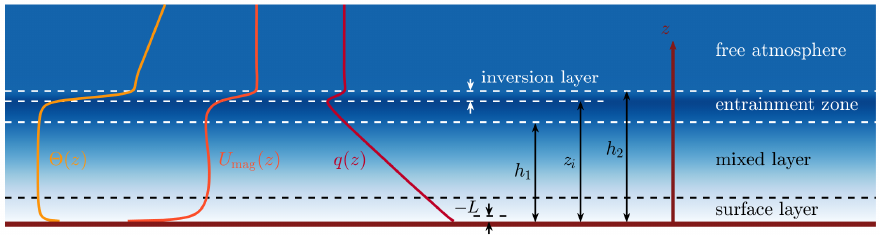}
\end{overpic}
\caption{Profiles of the potential temperature $\Theta$, wind speed $U_{\rm mag}$, and potential temperature flux $q$ in the CBL. The vertical lines with double arrows indicate different length scales in the CBL, namely, from left to right, the Obukhov length $L$, the lowest height where the potential temperature flux first reaches zero, $h_1$, the inversion layer height at which the potential temperature flux reaches its minimum value, $z_i$, and the height where the potential temperature flux first recovers zero, $h_2$. The background color indicates the magnitude of the potential temperature flux $q(z)$ for case 2, see Table~\ref{tab.summary}. }
\label{fig.sketch}
\end{figure*}

The CBL can be subdivided into three layers (excluding the roughness sublayer), i.e.\ the surface layer, the mixed layer, and the entrainment zone (see figure~\ref{fig.sketch}).\ The surface layer is characterized by a superadiabatic potential temperature gradient and a strong wind shear, which is usually described by the Monin-Obukhov similarity theory \citep[MOST,][]{mon54}. According to the MOST the non-dimensional wind speed and potential temperature gradient profiles are universal functions of the dimensionless height $z/L$, where $z$ is the height above the surface and $L$ is the surface Obukhov length \citep{obu46, mon54}. Many studies have pointed out that the MOST does not explain all important surface-layer statistics under convective conditions \citep{pan77, kha97, joh01, mcn07, sal20, che21} or very stable conditions \citep{mah98, che05}. In particular, the normalized wind gradient $\phi_m  = (\kappa z/u_*) (\pat U/\pat z)$ depends both on $z/L$ and $z_i/L$ \citep{kha97,joh01}, where $z_i$ is the height of the inversion layer (see figure~\ref{fig.sketch}).\ Nevertheless, MOST is still widely used in numerical weather prediction and climate models \citep{sal20}, and thus will be used in the theoretical analysis and numerical simulations of this study. MOST applies only to the surface layer, and for it to be applicable, the absolute value of the Obukhov length $L$ must be smaller than the height of the surface layer. Therefore we only consider the CBL with $-z_i / L \gg 1$. In particular, we focus on the convective-roll dominant regime with $-z_i/L\gtrsim10$ \citep{sal17}. Furthermore, we focus on dry and cloud-free CBLs to avoid complications due to physical processes like evaporation, precipitation, and cloud formation.

The mixed layer is characterized by intense vertical mixing caused by warm air thermals rising from the ground. Within the mixed layer, the magnitude of the mean velocity is much larger than the variations in the mean velocity. Thus, for applications where the mean velocity gradient is unimportant, the wind speed and potential temperature can be regarded as uniform \citep{kai76,sal17}. This insight is incorporated in various CBL models \citep{lil68, dea73, stu76, dea79, ten81, gar82}. The entrainment zone is characterized by entrainment of air from the free atmosphere. \citet{dea80} found in laboratory experiments that the ratio of the entrainment zone thickness to the depth of the mixing layer decreases asymptotically with increasing Richardson number $Ri$ as follows $(h_2-h_1)/h_1 = 0.21 + 1.31/Ri$ (see figure~\ref{fig.sketch} for the definitions of $h_1$ and $h_2$).\ This ratio is essential for developing entrainment models and has been studied extensively \citep{lil68, sul98, zil12, hag19}. The potential temperature flux profile decreases linearly with height and becomes negative in the entrainment zone.  The entrainment flux ratio $\Pi_m$, which is defined as the ratio of the potential temperature flux at the inversion layer height to its value at the ground, turns out to be nearly constant, i.e.\ $\Pi_m\approx-0.2$ \citep{stu76, sor96, con06-jas, sun08, lem19}. Note that the inversion layer is the upper region of the entrainment zone in which the potential temperature flux increases steeply from its minimum value at $z=z_i$ to zero at $z=h_2$ (see figure~\ref{fig.sketch}).

The geostrophic wind $(U_g, V_g)$ and the friction velocity $u_*$ are usually connected through the well-known geostrophic drag law, which was initially derived for neutral boundary layers \citep{ros35, bla68, ten72} and later extended to include buoyancy effects \citep{zil69}. To include the effect of unsteadiness, \citet{zil74} and \citet{ary75} proposed to replace the Ekman depth $u_*/|f|$ in the geostrophic draw law, where $f$ is the Coriolis parameter, with the time-dependent inversion layer height $z_i$. However, significant disparities were observed between the geostrophic drag law for CBLs and measurement data \citep{zil75}. \citet{gar82} derived a relationship for the velocity defects in the mixed layer using a three-layer CBL model, which accounts for the effects of entrainment, baroclinity, advection, and local acceleration.
In their formulation, the velocity defects are defined as the differences between the mixed-layer averaged winds and the geostrophic winds. They proposed a geostrophic drag law to relate the geostrophic winds and the friction velocity based on the assumption that the mean velocity at the top of the surface layer is equal to that in the mixed layer. In addition, an empirical stability function $\psi_m$, which may be inaccurate for large values of $-z/L$, is employed.

Recently, \citet{ton20} analytically derived the convective logarithmic friction law from first principles. They identified three scaling layers for the CBL with $-z_i / L \gg 1$: the outer layer, the inner-outer layer, and the inner-inner layer. The characteristic length scales for these three layers are the inversion layer height $z_i$, the Obukhov length $L$, and the roughness length $z_0$, respectively. The mixed-layer mean velocity scale $U_m$ and the geostrophic wind component $V_g$ are the characteristic streamwise and spanwise velocity scales in the outer layer. The difference between the horizontally and temporally averaged velocity $U(z)$ and $U_m$ is the mixed-layer velocity-defect law, which has a velocity scale of $u_*^2/w_* \ll U_m$ with $w_*$ the convective velocity. This indicates that $U_m$ is very close to the mean velocity $U(z)$ in the mixed layer. For the inner-outer layer they derived the surface-layer velocity-defect law, which states that the velocity defect $U-U_m$ scales with $u_*$. The convective logarithmic friction law is derived from matching the law of the wall in the inner-inner layer with the velocity-defect law in the surface-layer. This exact leading-order result relates the friction velocity $(u_*)$ to the mixed-layer velocity scale $(U_m)$. The difference between $U_g$ and $U_m$ scales as $(u_*^2 w_e) / (fz_i)^2$, where $w_e$ is the entrainment velocity and $V_g$ scales as $-u_*^2/(fz_i)$. Thus, up to non-dimensional coefficients, one can relate the geostrophic velocities $(U_g, V_g)$ to $u_*$. As \citet{ton20} do not consider the effects of the entrainment zone the velocity profiles are only valid for $z/z_i < 0.4$.

Various time-dependent models have been developed to explicitly account for entrainment processes at the top of CBLs \citep{tro86, noh03, hon06}. For example, the counter-gradient transport method \citep{hol91} and the eddy-diffusivity mass-flux approach \citep{sie07,li21} are widely used in coarse-resolution climate models. In general, the potential temperature is time-dependent \citep{lil68} and the entrainment velocity can affect the mean wind speed in the mixed layer \citep{ton20}. 
However, the velocity and potential temperature flux profiles are quasi-stationary, and therefore similarity theory can be employed to obtain analytical expressions for these profiles shapes \citep{zil74, ary75, zil92}. 

In this study, we focus on the the derivation of analytical expressions for the mean velocity and potential temperature flux profiles in cloud-free CBLs. We use a perturbation method approach to construct an analytical expression for the normalized potential temperature flux profile as a function of height, taking into account the characteristics of both the surface layer and the capping inversion layer.
The depth of the entrainment zone is connected to the convective logarithmic friction law to obtain analytical expressions for the velocity profile in the mixed layer and the entrainment zone. As remarked previously, the surface layer is still described by the MOST.

The organization of the paper is as follows. In Section~\ref{sec.theory} we obtain analytical expression for the potential temperature flux and wind profiles. In Section~\ref{sec.les} we validate the proposed profiles against LES. The conclusions are given in Section~\ref{sec.conclusions}.

\section{Theory}\lb{sec.theory}

\subsection{Potential temperature flux profile}

The potential temperature flux profile provides a precise and convenient demarcation between the mixed layer and the entrainment zone of the CBL \citep{dea79, dea80}. Figure~\ref{fig.sketch} shows a definition of the various length scales in the CBL. Previous studies \citep{kai76, dea80, moe94, noh03, gar14, hag19} showed that the potential temperature flux (including both the turbulent part and the diffusive part) in CBLs decreases linearly from its maximum value at the surface to a minimum value at $z=z_i$, and then increases steeply to zero in a narrow region $z_i \le z \le h_2$ at the top of the boundary layer (figure~\ref{fig.sketch}). For typical CBLs the condition $|\textrm{d} z_i/\textrm{d} t| \ll w_*$ holds, which implies that the boundary layer is quasi-stationary \citep[][Section 7.6]{nie16}. Besides, the potential temperature flux $q$ is fixed at the surface, and its value at the inversion layer height is nearly a constant fraction of the value at the ground $q_w$. Therefore, the normalized potential temperature flux $q(z,t)/q_w$ only depends on the similarity variable $\xi = z/h_2(t)$, i.e.
\beq\lb{eq.q3}
q(z,t)/q_w = \Pi (\xi), 
\eeq 
where the form of $\Pi$ remains to be determined. Using the potential temperature equation, we derive below an ordinary differential equation (ODE) for the determination of $\Pi$. However, it is important to emphasize that this does not mean $\Pi$ is independent of time as the similarity variable $\xi=z/h_2 (t)$ is still time-dependent.

Under the assumption of horizontal homogeneity, the potential temperature equation reduces to
\beq \lb{eq.theta}
\fr{\pat \Theta}{\pat t} = - \fr{\pat q}{\pat z} = -\fr{q_w}{h_2} \Pi',  
\eeq 
where $\Pi'=\pat \Pi/ \pat \xi$. In the mixed layer the potential temperature $\Theta$ is almost spatially uniform and hence the left-hand side of Eq.~\er{eq.theta} is independent of $z$. Therefore, the governing equation of the potential temperature flux in the mixed layer can be approximated as
\beq \lb{eq.q-1}
- \Pi' = c_\Pi,
\eeq 
where $c_\Pi$ is the gradient of the normalized potential temperature flux in the mixed layer (see figure~\ref{fig.sketch}). In the mixed layer, it is well-known that the eddy-diffusivity approach cannot adequately describe the potential temperature flux as the gradient of the potential temperature nearly vanishes \citep{wyn10}. In contrast, in the inversion layer the potential temperature gradient is dominant such that the potential temperature flux can be approximated by $q = - \nu_\theta \pat \Theta/\pat z$. Here $\nu_\theta \propto |w_e| (h_2-z_i)$ is the eddy diffusivity and $w_e=\textrm{d} z_i/ \textrm{d} t$ is the entrainment velocity. That the ratio $h_2/z_i$ is approximately constant implies that $\partial \xi/\partial t=-(z/h_2^2) (h_2/z_i) (\textrm{d} z_i/ \textrm{d} t) \approx -w_e/z_i$. From the zero-order jump model $\textrm{d} z_i^2/ \textrm{d} t$ is independent of time \citep[][Section 7.6]{nie16}, such that $\nu_\theta$ is approximately constant. Then, by taking the vertical derivative of Eq.~\er{eq.theta}, we obtain that 
\beq \lb{eq.q-2}
\epsilon \Pi'' - \Pi' = 0,
\eeq 
where $\epsilon$ is a small dimensionless parameter,
\beq \lb{eq.epsilon}
\epsilon \equiv c \fr{ h_2-z_i }{ h_2 }  \ll 1, 
\eeq 
with $c= (\nu_\theta z_i)/[-w_e h_2 (h_2-z_i)]=1/2$ being an empirical constant that is determined by comparing the model profiles to the simulation results. This indicates that the parameter $\epsilon$ represents the half thickness of the inversion layer normalized by the boundary layer depth (see figure~\ref{fig.sketch}).
To get the potential temperature flux profile in the entire boundary layer, we combine Eqs.~\er{eq.q-1} and \er{eq.q-2}, which leads to the following second-order ODE for the potential temperature flux, 
\beq\lb{eq.pi-ver1}
\epsilon \Pi'' - \Pi' = c_\Pi, \quad 
\Pi(0)=1, \quad 
\Pi(1)=0.
\eeq 

The solution of Eq.~\er{eq.pi-ver1} reads
\beqn\lb{eq.pi-final}
\Pi = 1 - c_\Pi \xi + (c_\Pi - 1) \frac{ e^{\xi/\epsilon } - 1 }{e^{1/\epsilon }-1}, \quad 
0\le \xi \le 1.
\eeqn
Since $\epsilon \ll 1$, the value of $\Pi$ in the bulk of the mixed layer can be approximated as 
\beqn\lb{eq.pi-ml}
\Pi \approx 1 - c_\Pi \xi.
\eeqn
Similarly, since $\Pi = 0$ at $\xi =h_1/h_2$, the slope $c_\Pi$ reduces to
\beqn\lb{eq.c-pi}
c_\Pi = h_2 / h_1 > 1.
\eeqn
Therefore, the ratio of the entrainment zone thickness to the mixing layer depth is
\beq \lb{eq.R}
R \equiv ({h_2-h_1})/{h_1} = c_\Pi - 1 > 0. 
\eeq 
\citet{dea80} found in laboratory experiments that the value of $R$ is between 0.2 and 0.4. Furthermore, the entrainment flux ratio $\Pi_m$, i.e.\ the minimum value of $\Pi$, can be approximated as
\beq 
\Pi_m \approx 1 - c_\Pi (1-2 \epsilon) \approx - (R-2\epsilon).\ 
\eeq 
\citet{stu76} and \citet{sor96} found that $-0.3 \le \Pi_m \le -0.1$. This results is consistent with the LES results of \citet{sul11} with $\Pi_m \approx -0.2$, the empirical results of \citet{len74} with $\Pi_m=-0.1$, and the direct numerical simulation results of \citet{gar14} with $\Pi_m \approx -0.12$.

We note that the perturbation method approach to model the potential temperature flux profile was recently introduced by \citet{liu21-prl} for conventionally neutral atmospheric boundary layers where the surface potential temperature flux is always zero. However, it should be noted that in the CBLs under consideration, the surface is heated and thermal plumes are generated at the ground, resulting in significantly different turbulence generation mechanisms. The applicability of the perturbation method approach to model the strong inversion layer relies on its ability to capture the strong gradients in the inversion layer. Here we used the second-order ODE defined by Eq.~\er{eq.pi-ver1} to model the potential temperature flux profile as our {\it a posteriori} tests confirm that this is sufficient to capture the inversion layer accurately. Higher-order terms could be incorporated, but this is not considered here to keep the obtained profiles relatively simple. An important observation is that the perturbation method approach is consistent with the finding of \citet{gar14}. They showed that the vertical structure of the entrainment zone is best described by two overlapping sublayers characterized by different length scales, namely the mean penetration depth of an overshooting thermal for the upper sublayer and the thickness of the CBL for the lower sublayer. Similarly, the second-order ODE, i.e. Eq.~\er{eq.pi-ver1}, indicates that there are two distinct length scales for the description of the entrainment zone (figure~\ref{fig.sketch}): one is the upper sublayer with $z_i\le z \le h_2$, where the gradient of potential temperature flux is proportional to $-(q_w \Pi_m)/(2 \ep h_2)$, and the other is the lower sublayer with $h_1\le z \le z_i$, where the gradient of potential temperature flux is proportional to $(q_w \Pi_m)/z_i$. Since $\ep \ll 1$ the potential temperature flux varies much steeper in the upper sublayer than in the lower sublayer.

\subsection{Wind profile}

We consider MOST to describe the wind speed profile in the surface layer \citep{mon54}. In surface-layer coordinates, it states that the non-dimensional streamwise velocity gradient can be written as
\beq \lb{eq.dU-sl}
\fr{\kappa z}{u_*} \fr{ {\textrm d} U}{{\textrm d} z} = \phi_m \left( \fr{z}{L} \right),
\eeq 
where $\phi_m$ is the dimensionless stability correction function and $L = - u_*^3 / (\ka \beta q_w)$ is the Obukhov length with $\beta$ the buoyancy parameter. By integrating Eq.~\er{eq.dU-sl}, one can obtain the explicit formula for the streamwise velocity $U$,
\beq \lb{eq.U-sl}
\fr{\kappa U }{u_*} = \ln \left( \fr{z}{z_0} \right) - \psi_m \left( \fr{z}{L} \right).
\eeq 
Here $\kappa=0.4$ is the von K\'arm\'an constant, $z_0$ is the roughness length, and 
\beq\lb{eq.psi}
\psi_m = \int_{z_0/L}^{z/L} \fr{ 1-\phi_m(\zeta)}{\zeta} {\textrm d} \zeta
\eeq
is the stability correction function for the momentum. We use the well-known Businger-Dyer expression \citep{pau70, bus71, dye74, bru82}
\beq\lb{eq.psim}
\psi_m = \ln \fr{(1+x^2)(1+x)^2}{8} - 2 \arctan x + \fr{\pi}{2}, 
\eeq
with $x = (1-16z/L)^{1/4}$ to model the stability correction function, but we note that other parameterizations exist \citep{kat11}. Note that $\psi_m=0$ for $x=1$ (or $L=\infty$), which reduces Eq.~\er{eq.U-sl} to the classical logarithmic law for neutral boundary layers.

To model the wind profile $U$ in the mixed layer and entrainment zone, we again employ a second-order ODE. In general, the detailed wind profile evolves when stability changes. However, the variations of the mean velocity are small compared to the magnitude of the mean velocity in the mixed layer when the stability parameter $-z_i/L \gtrsim 10$ \citep[e.g.][]{sal17}, which covers the range of stability conditions considered in this study. Thus, we can assume that the ODE is dominated by the $U' = 0$ term in most of the domain. Recently, \citet{ton20} derived the convective logarithmic friction law from first principles, which connects the mixed-layer mean velocity scale $U_{m}$ and the friction velocity $u_*$ in the convective-roll dominant regime ($-z_i/L \gg 1$) as follows,
\beq \lb{eq.um}
\fr{ U_m }{u_*} = \fr{1}{\kappa} \ln \left(- \fr{L}{z_0} \right) - C,
\eeq 
where $C=1$ is an empirical constant determined from our LES database (see below). 

From the potential temperature flux profile modeling we learned that the ODE should have a second-order derivative term $\epsilon U''$ to model the entrainment zone near the top of the boundary layer. The top boundary condition is given by the geostrophic wind component $U_g$. The lower boundary condition is given by equaling Eqs.~\er{eq.U-sl} and \er{eq.um}, namely $U (\xi_0) = U_m$, since \citet{ton20} showed that $U_m$ is very close to $U(z)$ in the mixed layer. Here $\xi_0$ represents the height of the top of surface layer, which can be determined using Eqs.~\er{eq.U-sl} and \er{eq.um},
\beq \lb{eq.xi0}
 \ln \left( -\fr{h_2}{L} \xi_0 \right) - \psi_m \left( \fr{h_2}{L} \xi_0 \right) = - \kappa C \quad \Rightarrow \quad \xi_0 = \xi_0\left(\fr{h_2}{L} \right).
\eeq 
Because $\ep \ll 1$, the solution obtained from $U(\xi_0)=U_m$ is almost the same as from $U (0)=U_m$, while the expression of the latter is much simpler. Therefore, we model the profile of the streamwise velocity $U$ in the mixed layer and the entrainment zone as:
\beq\lb{eq.U-ver1}
\epsilon U'' - U' = 0, \quad 
U (0)=U_m, \quad 
U (1)=U_g.
\eeq 
The solution of Eq.~\er{eq.U-ver1} is 
\beq\lb{eq.U-cl}
U = U_m +(U_g-U_m) \frac{e^{\xi/\epsilon }-1}{e^{1/\epsilon }-1}.
\eeq 
We note that Eq.~\er{eq.U-cl} is only valid in the mixed layer and entrainment zone as the wind speed in the surface layer is still modeled using the MOST. By combining Eqs.~\er{eq.U-sl} and \er{eq.U-cl} and recalling that Eq.~\er{eq.U-sl} increases monotonically as $z$ increases, we obtain the following analytic description of the streamwise velocity profile $U(z)$ for the entire CBL,
\beq\lb{eq.U-final}
U = \left\{
\begin{split}
    & \fr{u_*}{\kappa} \left[ \ln \left( \fr{z}{z_0} \right) - \psi_m \left( \fr{z}{L} \right) \right], \quad & & \xi\le\xi_0, \\
    & U_m +(U_g-U_m) \frac{e^{\xi/\epsilon }-1}{e^{1/\epsilon }-1}, \quad & & \xi_0<\xi \le 1,
\end{split}
\right.
\eeq 
where $\xi_0$ is given by Eq.~\er{eq.xi0}.
As remarked in Section~\ref{sec.introduction} the surface layer profile contains two length scales, i.e.\ $z_0$ for the inner-inner layer and the Obukhov length $L$ for the inner-outer layer. Similarly, the velocity profile for the mixed layer and entrainment zone contains two length scales, i.e.\ $z_i$ to describe the mixed layer and $h_2-z_i=2\ep h_2$ to describe the upper sublayer of the entrainment zone. This confirms the view presented by \citet{ton20} that the entrainment zone has a different scaling than the surface and mixed layers, and can therefore be considered as another inner layer in the overall CBL problem. We note that the proposed analytical profile is empirical, similar to the MOST, and that the parameter $\ep$ parameterizes the effect of various physical processes. We further note that $U_m$ and $u_*$ are related as given by Eq.~\er{eq.um}, and that the difference $U_g-U_m$ scales as $(u_*^2 w_e) / (fz_i)^2$ \citep{ton20}. Thus, Eq.~\er{eq.U-final} is predictive if the entrainment velocity $w_e$ is given as an input parameter. To determine the value of $w_e$, one may need to revisit the entrainment processes at the top of CBLs \citep[e.g.][]{gar14}. In addition, the velocity predicted by Eq.~\er{eq.U-final} is continuous throughout the boundary layer and applicable for the considered ranges (see figure~\ref{fig.umag} below). However, its first derivative is discontinuous at the patching location $\xi=\xi_0$. This is a typical character of low-order models \citep{gar82}. To capture the smooth transition, a high-order model is needed \citep{ton20}.
We leave these for future work.

To model the wind profile $V$ in the the mixed layer and entrainment zone, we use a similar ODE as Eq.~\er{eq.U-ver1}. The top boundary condition is given by the geostrophic wind component $V_g$. As the spanwise velocity $V$ is small compared to the streamwise velocity $U$ in the mixed layer \citep{ton20}, the lower boundary condition is given by $V(\xi_0)= V(0)=0$. Therefore, we model the profile of the spanwise velocity $V$ in the entire boundary layer using
\beq\lb{eq.V-ver1}
\epsilon V'' - V' = 0, \quad 
V (0)=0, \quad 
V (1)=V_g.
\eeq 
The solution of Eq.~\er{eq.V-ver1} is 
\beq\lb{eq.V-final}
V = V_g \frac{e^{\xi/\epsilon }-1}{e^{1/\epsilon }-1}.
\eeq 
Since $V_g$ scales as $-u_*^2/(fz_i)$ \citep[e.g.][]{wyn10, ton20}, the geostrophic wind component $V_g$ can be connected to $u_*$, up to a non-dimensional coefficient $-V_g fz_i/u_*^2=0.66$, which is determined from our LES database (see Table~\ref{tab.summary}).

\section{Numerical validation}\lb{sec.les}

\subsection{Numerical method and computational setup}

We use LES to simulate the CBL flow over an infinite flat surface with homogeneous roughness. We integrate the spatially-filtered Navier-Stokes equations and the filtered transport equation for the potential temperature \citep{alb96, alb99, gad21, liu21-qjrms, liu21-prl, liu21-re}. Molecular viscosity is neglected as the Reynolds number in the atmospheric boundary layer flow is very high, and we use the advanced Lagrangian-averaging scale-dependent model to parameterize the sub-grid scale shear stress and potential temperature flux \citep{bou05, sto08}. We note that the Lagrangian-averaging scale-dependent model has been extensively validated and widely used in the literature \citep{bou05, sto08, cal10, wu11, zha19, gad21}.

Our code is an updated version of the one used by \citet{alb99}. The grid points are uniformly distributed, and the computational planes for horizontal and vertical velocities are staggered in the vertical direction. The first vertical velocity grid plane is located at the ground. The first gridpoint for the horizontal velocity components and the potential temperature is located at half a grid distance above the ground. We use a second-order finite difference method in the vertical direction and a pseudo-spectral discretization in the horizontal directions. Time integration is performed using the second-order Adams-Bashforth method. The projection method is used to enforce the divergence-free condition. At the top boundary, we impose a constant potential temperature lapse rate, zero vertical velocity, and zero shear stress boundary condition. At the bottom boundary, we employ the classical wall stress and potential temperature flux formulations based on the MOST \citep{moe84, bou05, sto08, gad21}. 

We perform eleven LES to verify the validity of the derived wind speed and potential temperature flux profiles for CBLs. The computational domain is $5 \, \textrm{km} \times 5 \, \textrm{km} \times 2 \, \textrm{km}$ and the grid resolution is $256 \times 256 \times 256$. Due to large computational expense, only several external parameters are varied in the simulations. The flow is driven by the geostrophic wind of $G=\sqrt{U_g^2+V_g^2} = 10$~m/s, the buoyancy parameter is $\beta =0.0325 \, \textrm{m}/ (\textrm{s}^2 \cdot \textrm{K})$, and the Coriolis parameter is $f = 1 \times 10^{-4} \, \textrm{rad/s}$ \citep{moe94, abk17, gad21}. To ensure the CBLs are in the convective-roll dominant regime with $-z_i/L \gtrsim 10$, the surface potential temperature flux is set to $q_w = 0.12 \sim 0.24 \, \textrm{K} \cdot \textrm{m/s}$. Note that the convective logarithmic friction law (Eq.~\er{eq.um}) is derived very recently by \citet{ton20} and tested only in a relatively narrow range of $-L/z_0$, namely $-L/z_0 \in [2.5\times10^2, 1.5 \times 10^3]$. To evaluate the performance of this law in much wider range, i.e. $-L/z_0 \in [3.6\times10^2, 0.7 \times 10^5]$, the roughness length is varied between $z_0 = 0.0002 ~\textrm{m}$ and $z_0 = 0.16~ \textrm{m}$, where the lower bound of $z_0$ is set to a representative value of the sea surface \citep{wie01}. The vertical potential temperature gradient is varied between $\Gamma=1$~K/km and $\Gamma=9$~K/km to capture the relevant range observed in atmospheric measurements \citep{sor96}. The velocity field is initialized with the geostrophic wind $G=10$~m/s. The initial potential temperature is 300~K up to 937~m and increases with 8~K in the next 126~m above. Above $1063$~m the constant vertical derivative of the potential temperature $\Gamma$ is specified. The simulations are run for about 25 large-eddy turnover times $T=z_i/w_*$, where $w_* = (\beta q_w z_i)^{1/3}$ is the convective velocity scale, and the statistics are computed from the time interval of $12T$ to $25T$ when the boundary layer is quasi-stationary \citep{din21}. We note that inertial oscillation develops as the flow is initialized with a profile in geostrophic equilibrium \citep{sch13}. However, it should have only negligible effect on the simulated statistics since the typical large-eddy turnover time $T\approx10$~mins is two order smaller than the natural inertial periodicity $2\pi/f \approx 17.5$~h.

\begin{figure*}[!tb]
\centering
\begin{overpic}[width=0.75\textwidth]{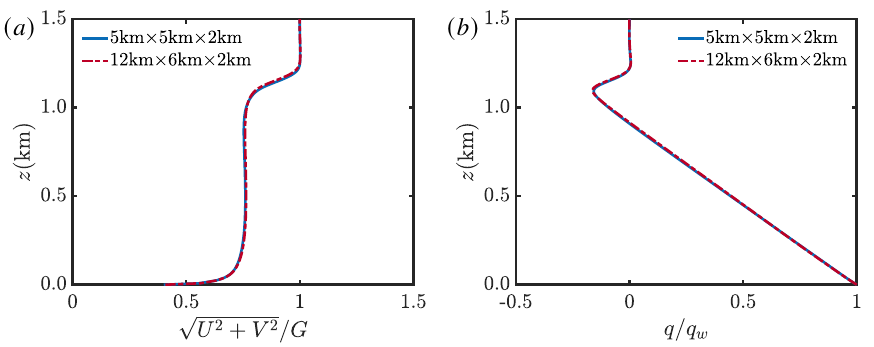}
\end{overpic}
\caption{The comparison of simulated (a) normalized wind speed $\sqrt{U^2+V^2}/G$ and (b) normalized potential temperature flux $q/q_w$ profiles for case 2 (see Table~\ref{tab.summary}) with different computational domain sizes.}
\label{fig.domain}
\end{figure*}

A summary of all simulated cases is presented in Table~\ref{tab.summary}. Note that the cases in Table~\ref{tab.summary} are arranged such that the value of $-L/z_0$ increases monotonically. Furthermore, we note that case 2 has been validated against atmospheric observations, and the simulation results obtained using different sub-grid scale models and grid resolutions is very similar \citep{gad21}. To show the simulated results are independent of the computational domain size, we have performed an additional simulation for case 2 in a larger computational domain ($12 \, \textrm{km} \times 6 \, \textrm{km} \times 2 \, \textrm{km}$) on a mesh with $600 \times 300 \times 240$ nodes such that the grid spacings are nearly identical. Figure~\ref{fig.domain} shows the simulated wind speed and potential temperature flux profiles for case 2. The good agreement between the results  obtained with different computational domain size confirms that the simulated results are independent of the computational domain size.

\begin{table*}[!tb]
\caption{Summary of all simulated cases. Here $\Gamma=\pat \Theta / \pat z$ is the vertical derivative of the mean potential temperature in the free atmosphere, $q_w$ is the surface potential temperature flux, $z_0$ is the roughness length, $u_*$ is the friction velocity, $U_m$ is the wind speed in the mixed layer, $L$ is the Obukhov length, $z_i$ is the inversion layer height, $|V_g|$ is the magnitude of the spanwise geostrophic wind, $\ep$ and $c_\Pi$ are dimensionless parameters calculated by Eqs.~\er{eq.epsilon} and \er{eq.c-pi}, respectively, and $Ri=\beta \Delta \Theta z_i / w_*^2$ is the Richardson number, where $\Delta \Theta = \Theta(h_2)-\Theta(h_1)$ is the potential temperature difference across the entrainment zone and $w_* = (\beta q_w z_i)^{1/3}$ is the convective velocity.}
\lb{tab.summary}
\begin{center}
\begin{tabular}{cccccccccccc}
\topline
 Case & $\Gamma$ (K/km) & $q_w$ (K$\cdot$m/s) & $z_0$ (m) & $u_*$ (m/s) & $U_m$ (m/s) & $|V_g|$ (m/s) & $\epsilon$ & $c_\Pi$ & $Ri$ & $-z_i/L$ & $-L/z_0$ \\
\midline
1 & 9 & 0.24 & 0.16 & 0.562 & 7.60 & 2.00 & 0.044 & 1.32 & 56.1 & 19.2 & $3.6\times10^2$ \\
2 & 3 & 0.24 & 0.16 & 0.563 & 7.59 & 1.87 & 0.052 & 1.34 & 51.0 & 19.1 & $3.6\times10^2$ \\
3 & 1 & 0.24 & 0.16 & 0.562 & 7.59 & 1.84 & 0.055 & 1.34 & 47.9 & 19.2 & $3.6\times10^2$ \\
4 & 3 & 0.12 & 0.16 & 0.533 & 7.70 & 2.20 & 0.046 & 1.34 & 94.2 & 11.0 & $0.6\times10^3$ \\
5 & 3 & 0.24 & 0.016 & 0.463 & 8.44 & 1.17 & 0.050 & 1.33 & 51.0 & 34.5 & $2.0\times10^3$ \\
6 & 3 & 0.20 & 0.02 & 0.468 & 8.36 & 1.32 & 0.046 & 1.31 & 59.6 & 27.5 & $2.0\times10^3$ \\
7 & 3 & 0.12 & 0.016 & 0.444 & 8.45 & 1.43 & 0.038 & 1.31 & 91.5 & 19.0 & $3.5\times10^3$ \\
8 & 3 & 0.20 & 0.002 & 0.392 & 8.93 & 0.86 & 0.033 & 1.28 & 55.5 & 47.3 & $1.2\times10^4$ \\
9 & 3 & 0.24 & 0.0016 & 0.389 & 8.94 & 0.75 & 0.044 & 1.30 & 50.4 & 58.3 & $1.2\times10^4$ \\
10 & 3 & 0.12 & 0.0016 & 0.375 & 8.94 & 0.88 & 0.036 & 1.30 & 93.9 & 31.6 & $2.1\times10^4$ \\
11 & 3 & 0.20 & 0.0002 & 0.334 & 9.24 & 0.57 & 0.041 & 1.30 & 58.6 & 75.8 & $0.7\times10^5$ \\
\botline
\end{tabular}
\end{center}
\end{table*}

\subsection{Validation of analytical profiles}

\begin{figure*}[!tb]
\centering
\begin{overpic}[width=0.5\textwidth]{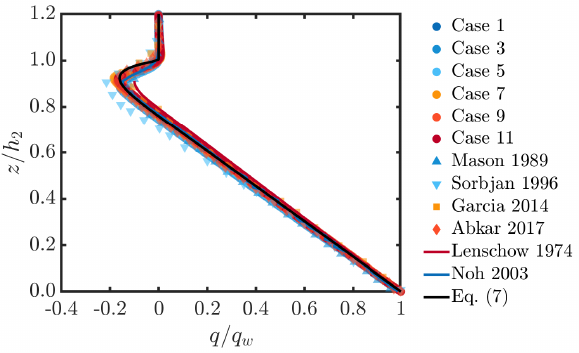}
\end{overpic}
\caption{Vertical profile of normalized potential temperature flux $q / q_w$. Circles:\ LES data (Table~\ref{tab.summary}); up-triangle:\ LES data of \citet{mas89}; down-triangle:\ LES data of \citet{sor96}; square:\ direct numerical simulations data of \citet{gar14}; diamond:\ LES data of \citet{abk17}; red line:\ prediction given by \citet{len74}; blue line:\ prediction given by \citet{noh03}; black line:\ prediction given by Eq.~\er{eq.pi-final} with $\ep=0.044$ and $c_\Pi=1.32$. The figure shows that the proposed model captures the simulation trends very well.}
\label{fig.q}
\end{figure*}

Figure~\ref{fig.q} shows that with increasing height the normalized potential temperature flux profile $q / q_w$ first decreases linearly from unity at the surface to a minimum at $z_i/h_2 = 1-2\ep$, before it rapidly increases to zero in the inversion layer ($1-2\epsilon \le z/h_2 \le 1$).\ The normalized thickness of the inversion layer, which is parameterized by $\epsilon$, is expected to depend on the Richardson number \citep{dea80}, potential temperature gradient \citep{sor96}, and wind shear \citep{con06-jas}. However, we find that for the parameter range under consideration, the variation in the normalized thickness of the inversion layer is limited (see Table~\ref{tab.summary}). Therefore, we use a fixed representative value $\epsilon=0.044$ to model the potential temperature flux profile, and the figure confirms that this ensures that the potential temperature flux profile obtained from the model agrees excellently with all available simulation data, which validates the chosen approach. To further confirm the validity of the potential temperature flux profile, we also compare our results in figure~\ref{fig.q} with previous LES from \citet{mas89}, \citet{sor96}, and \citet{abk17}, the direct numerical simulations data by \citet{gar14}, and the empirical models by \citet{len74} and \citet{noh03}. Clearly, the model predictions agree well with these previous studies.

\begin{figure*}[!tb]
\centering
\begin{overpic}[width=0.75\textwidth]{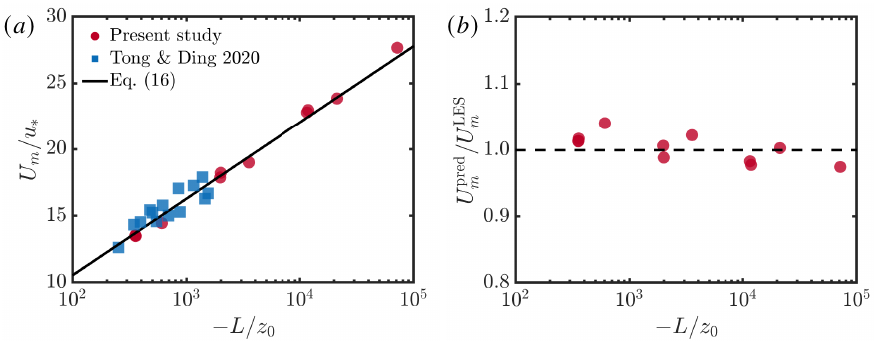}
\end{overpic}
\caption{The dependence of {(a) the normalized mixed-layer mean velocity scale $U_m/u_*$ and (b) the ratio of the predicted and simulated mixed-layer mean velocity scale $U_m^{\rm pred}/U_m^{\rm LES}$ against the normalized Obukhov length} $-L/z_0$. Circles:\ LES data (Table~\ref{tab.summary}); squares: LES data of \citet{ton20}; solid line:\ prediction given by Eq.~\er{eq.um} with $C=1$ determined using a fit to the present simulation data. }
\label{fig.Um}
\end{figure*}

\begin{figure*}[!tb]
\centering
\begin{overpic}[width=0.75\textwidth]{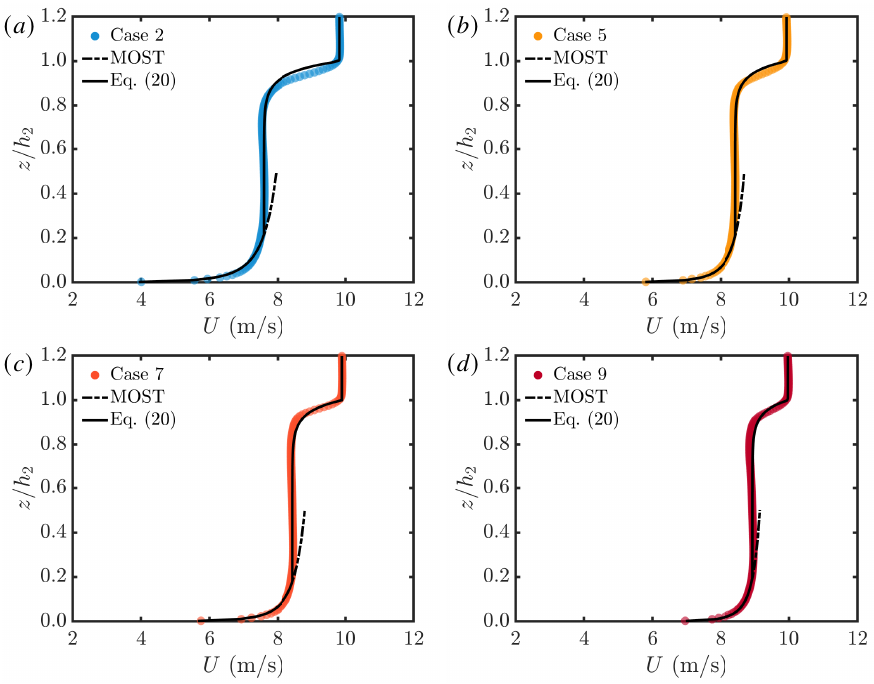}
\end{overpic}
\caption{Vertical profile of the mean streamwise velocity $U$. Filled circles:\ LES data (Table~\ref{tab.summary}); dashed line:\ prediction given by the MOST; solid line:\ prediction given by Eq.~\er{eq.U-final} with $\epsilon=0.044$. The prediction by the MOST is plotted outside the surface layer region to demonstrate the difference with the new profile. }
\label{fig.umag}
\end{figure*}

In figure~\ref{fig.Um}(a) we compare the wind speed in the bulk of the mixed layer ($0.4 \le z/h_2 \le 0.6$) against the normalized Obukhov length $-L/z_0$ with results from \citet{ton20}. The figure shows that our simulations convincingly confirm the validity of the convective logarithmic friction law for the wind speed (Eq.~\er{eq.um} with $C=1$) over a much wider range of $-L/z_0 \in [3.6\times10^2, 0.7 \times 10^5]$ than previously considered ($-L/z_0 \in [2.5\times10^2, 1.5 \times 10^3]$). To further confirm the convective logarithmic friction law, figure~\ref{fig.Um}(b) shows the ratio of the predicted and simulated mixed-layer mean velocity scale $U_m^{\rm pred}/U_m^{\rm LES}$ against the normalized Obukhov length $-L/z_0$. The figure shows that the predicted value is within 5\% of the simulation result for all considered cases.

\begin{figure*}[!tb]
\centering
\begin{overpic}[width=0.75\textwidth]{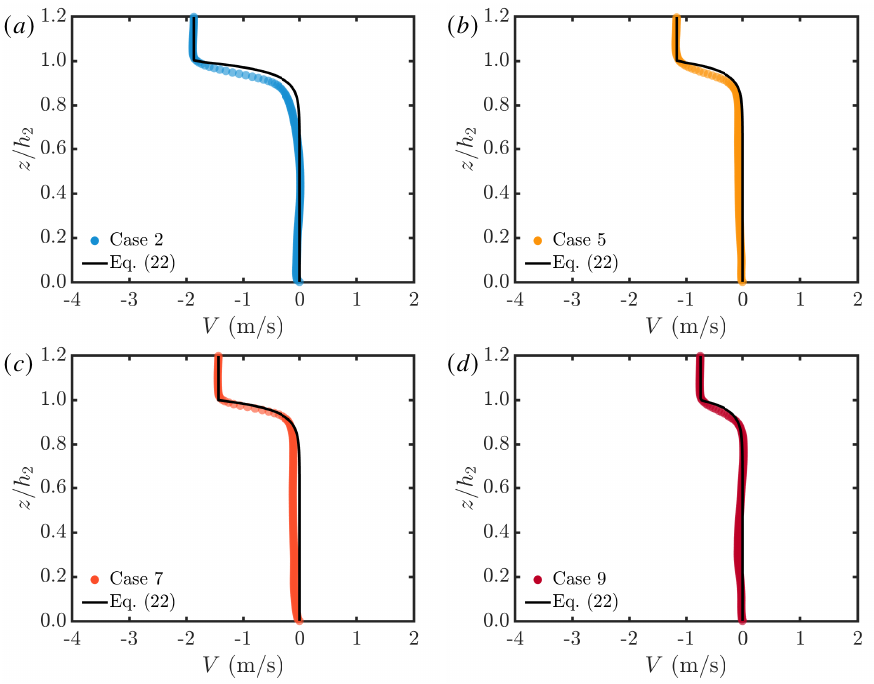}
\end{overpic}
\caption{Vertical profile of the mean spanwise velocity $V$. Filled circles:\ LES data (Table~\ref{tab.summary}); solid line:\ prediction given by Eq.~\er{eq.V-final} with $\epsilon=0.044$. }
\label{fig.V}
\end{figure*}

In figure~\ref{fig.umag} we compare the vertical profile of the mean streamwise velocity $U$ for four typical cases with different surface potential temperature flux and roughness length with the simulation results. The filled symbols are the present LES data, the dashed line is the theoretical prediction given by the MOST, and the solid line is the prediction given by Eq.~\er{eq.U-final} with $\epsilon=0.044$. The figure shows that the MOST accurately captures the surface layer's wind profile (lowest $20\%$ of the boundary layer).\ However, in the mixed layer, the prediction of the MOST deviates significantly from the LES data. In particular, the discrepancy from the MOST increases as the surface potential temperature flux $q_w$ decreases (figure~\ref{fig.umag}b,c) or the roughness length $z_0$ increases (figure~\ref{fig.umag}a,d).\ Therefore, MOST is seldom used to specify wind profiles outside of the surface layer. Figure~\ref{fig.umag} shows that the proposed wind profile given by Eq.~\er{eq.U-final} accurately captures the velocity profile throughout the entire boundary layer. This excellent agreement confirms the validity of our proposed wind profile of Eq.~\er{eq.U-final} for atmospheric boundary layers in the range of studied parameters.

Figure~\ref{fig.V} shows the corresponding profiles of the mean spanwise velocity $V$. The filled symbols are the present LES data and the solid line is the prediction given by Eq.~\er{eq.V-final} with $\epsilon=0.044$. Overall, the agreement between the proposed wind profile given by Eq.~\er{eq.V-final} and the LES data is reasonably good in the entire boundary layer. This agreement confirms the validity of our proposed wind profile of Eq.~\er{eq.V-final} for CBLs in the range of studied parameters (i.e. $-L/z_0 \in [3.6\times10^2, 0.7 \times 10^5]$). We note that the figure confirms that the spanwise velocity $V$ is much smaller than the streamwise velocity $U$. The figure also indicates that the magnitude of the geostrophic wind component $|V_g|$ increases as the surface potential temperature flux $q_w$ decreases (figure~\ref{fig.V}b,c) or the roughness length $z_0$ increases (figure~\ref{fig.V}a,d).

\section{Conclusions}\label{sec.conclusions}

This work uses a perturbation method approach in conjuncture with the convective logarithmic friction law and the Monin-Obukhov similarity theory to develop analytical expressions of the wind and potential temperature flux profiles in convective atmospheric boundary layers. The validity of the proposed wind (given by Eqs.~\er{eq.U-final} and \er{eq.V-final}) and potential temperature flux profiles (given by Eq.~\er{eq.pi-final}) has been confirmed by their excellent agreement with large-eddy simulations results for atmospheric boundary layers in the convective-roll dominant regime with $-z_i/L \gtrsim 10$, where $L$ is the Obukhov length and $z_i$ the inversion layer height. Furthermore, our simulations confirm that the convective logarithmic friction law of Eq.~\er{eq.um}, which was originally proposed by \citet{ton20} for the mixed-layer mean velocity scale, is valid for an extensive range of $-L/z_0$, namely $-L/z_0 \in [3.6\times10^2, 0.7 \times 10^5]$, where $z_0$ is the surface roughness length. Since accurate capturing the coupling between meso and microscale processes is a long-standing challenge in numerical weather predictions \citep{wyn04, lar18, vee19}, the proposed analytical profiles may be relevant for climate modeling and weather forecasting to better understand the effect of convective atmospheric boundary layers on, for example, wind farms. Possible future work will involve investigating models to predict the entrainment velocity at the top of CBLs and developing a high-order model that can capture the transition between the entrainment zone and free atmosphere. The latter may require a formal asymptotic series expansion of the governing equations, allowing for the separation into a time-dependent and steady-state problem at different orders.

\acknowledgments
This work was supported by the Hundred Talents Program of the Chinese Academy of Sciences, the National Natural Science Fund for Excellent Young Scientists Fund Program (Overseas), the National Natural Science Foundation of China Grant (No. 11621202), the Shell-NWO/FOM-initiative Computational sciences for energy research of Shell and Chemical Sciences, Earth and Live Sciences, Physical Sciences, Stichting voor Fundamenteel Onderzoek der Materie (FOM) and STW, and an STW VIDI Grant (No. 14868).\ This work was sponsored by NWO Domain Science for the use of the national computer facilities. We acknowledge PRACE for awarding us access to Irene at Tr\`es Grand Centre de Calcul du CEA (TGCC) under PRACE project 2019215098, and the advanced computing resources provided by the Supercomputing Center of the USTC.

%
%
\datastatement
The data that support the findings of this study are available from the corresponding author upon reasonable request.

\bibliographystyle{ametsocV6}
\bibliography{windfarm}

\end{document}